\documentstyle[prl,aps,epsfig]{revtex}

\makeatletter 
\renewcommand{\ps@plain}{%
  \renewcommand{\@oddfoot}{\hfil\ Phys.\ Rev.\ Lett. {\bf 83} (1999)
    2872--2875 \qquad \qquad \thepage \qquad \qquad \copyright 1999
    The American Physical Society \hfil}}
\makeatother

\renewcommand{\L}{\text{L}}
\newcommand{\E}{\text{E}}
\renewcommand{\d}{\text{d}}
\newcommand{\ret}{\text{ret}}
\newcommand{\El}{E_{\L\L}}
\newcommand{\Ee}{E_{\E\E}}
\newcommand{\Es}{E_{\text{coinc.}}}

\newcommand{\parsign}{\mathchar "278}
\twocolumn

\begin{document}
\title{Two-photon Franson-type experiments and local realism}
 
\author{Sven Aerts$^{1,*} $, Paul Kwiat$^{2 ,\dagger }$, Jan-{\AA}ke
  Larsson$^{3 ,\ddagger }$, and Marek \.Zukowski $^{4 ,\parsign } $ }
\address{$^1$Fundamenten van de Exacte Wetenschappen, Vrije
  Universiteit Brussel, Triomflaan 2, 1050 Brussel, Belgium}
\address{$^2$ P-23, MS-H803, Los Alamos National Laboratory, Los
  Alamos 87545, USA} \address{$^3$Matematiska Institutionen,
  Link\"opings Universitet, SE-581 83 Link\"oping, Sweden}
\address{$^4$Instytut Fizyki Teoretycznej i Astrofizyki Uniwersytet
  Gda\'nski, PL-80-952 Gda\'nsk, Poland} \date{27 May 1999}

\maketitle
\widetext
\begin{quote}
  \begin{abstract}
    The two-photon interferometric experiment proposed by Franson
    [Phys.\ Rev.\ Lett.\ 62, 2205 (1989)] is often treated as a ``Bell
    test of local realism''. However, it has been suggested that this
    is incorrect due to the $50\%$ postselection performed even in the
    ideal gedanken version of the experiment. Here we present a simple
    local hidden variable model of the experiment that successfully
    explains the results obtained in usual realizations of the
    experiment, even with perfect detectors.  Furthermore, we also
    show that there is \emph{no} such model if the switching of the
    local phase settings is done at a rate determined by the
    \emph{internal geometry of the interferometers}.
  \end{abstract}  
\end{quote}
\narrowtext

\vspace*{-.7cm}
The two-particle interferometer introduced by Franson \cite{FRANSON89}
has been used in many two-photon interferometric experiments
\cite{FR-EXP,CRYPTO} that reveal complementarity between single and
two-photon interference.  The experiments cannot be described using
standard methods involving classical electromagnetic fields
\cite{CLASSICAL}. However, the original paper was entitled \emph{Bell
  Inequality for Position and Time}, and many subsequent papers
claimed that the experiment constitutes a ``Bell test of local realism
involving time and energy''. Some authors were more skeptical that a
true, unambiguous test of a Bell inequality was possible with these
experiments, even in principle, since even the ideal gedanken model of
the experiment requires a post-selection procedure in which $50\%$ of
the events are discarded when computing the correlation functions
\cite{GARUCCIO,KWIAT2}. If all events are taken into account the Bell
inequalities are not violated. Thus, a local hidden-variable (LHV)
model is not ruled out, but even so, no LHV model for the experiment
has yet been constructed \cite{SANTOS}.

The situation is further obscured by similar claims concerning certain
other two-photon polarization experiments \cite{MANDELSHIH} where the
problem of discarding $50\%$ of the events also appears
\cite{GARUCCIO,KWIAT}. This was initially treated on equal footing
with the problems of Franson-type experiments, but a recent analysis
in \cite{POPESCU} reestablishes the possibility of violating local
realism. Unfortunately, that analysis cannot be adapted to the Franson
experiment.

Our aim is to resolve this uncertainty. First, we shall construct a
simple local realistic model for the usual operational realization of
the experiment. Second, we shall prove that under the additional
condition that the random changes of the state of the local
interferometers are at a rate dictated by the \emph{internal geometry}
of the interferometers, \emph{no} local hidden variable model exists
for the perfect gedanken version of this type of experiment. Even
then, the usual Bell inequality will be inadequate.\\

\vspace*{2.1cm}

\begin{figure}[htbp]
  \begin{center}
    \psfig{file=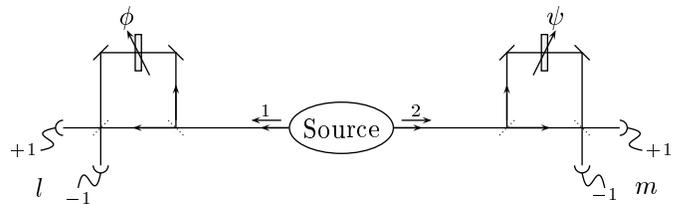}
    \caption{The generic setup of the
      Franson two-photon interference experiment. }
    \label{fig:exp}
  \end{center}
\end{figure}

Let us briefly describe the idea behind the Franson-type experiments
(Fig.~\ref{fig:exp}).  The source yields photon pairs, correlated to
within their coherence times, and the two photons are fed into two
identical unbalanced Mach-Zehnder interferometers. The difference of
the optical paths in those interferometers, $\Delta L$, satisfies the
relation $ \Delta L \gg cT_{\text{coh}} $, where $c$ is the speed of
light and $T_{\text{coh}}$ is the coherence time of the photons. Such
optical path differences prohibit any single-photon interference, so
the single-photon probabilities are $\textstyle
P(l|\phi)=P(m|\psi)=\frac{1}{2}$ (see Fig.~\ref{fig:exp}). For the
$50\%$ two-photon events that are coincident, one cannot distinguish
between events where both photons take the long path and events where
both take the short; hence, two-photon interference occurs:
\begin{equation}
  P\bigl(l;m(\text{coinc.})|\phi,\psi\bigr)
  =\textstyle\frac{1}{8}\bigl(1+lm\cos(\phi+\psi)\bigr).
  \label{eq:coinc}
\end{equation}
For the other half of the two-photon events, one photon takes its
short path and the other takes its long path, so that the registration
times differ by $\Delta L/c$; there is consequently no interference
because the events are distinguishable. One has $
P(l_{\L};m_{\E}|\phi,\psi) =P(l_{\E};m_{\L}|\phi,\psi)
=\textstyle\frac{1}{16}$, where E denotes the \emph{earlier} count,
and L denotes the \emph{later} count.  For future reference, we note
that the local phase settings appearing in these formulas are those
present when a photon in the long path is passing through the phase
shifter, i.e., the phase setting at the actual detection time
$t_{\d}$, minus the time $t_{\ret}$ it takes light to reach the
detector from the location of the phase-shifter by the optical paths
available within the interferometer.

Initially, the experiment is assumed to be performed in the following
way. The usual locality condition is imposed, i.e., the local phase
setting at one side does not affect the measurement result at the
other side. Experimentally, this is enforced by switching the local
phase settings on the time-scale $D/c$, where $D$ is the
source--interferometer distance. We assume that $D\gg\Delta L$
\cite{AD-HOC}.  The two experimenters (one at each side) record the
$\pm 1$ counts, the detection times, and the appropriate values of the
local phase settings. After the experiment is completed they perform a
post-detection analysis on their recorded data, rejecting all pairs of
events whose registration times differ by $\Delta L/c$. We now present
a LHV model for the Franson experiment, valid in this experimental
situation.

There are some general features that a LHV model of the experiment
should have. The \emph{emission time} should be one of the variables,
because if the beamsplitters of, say, the right interferometer were
removed, the photons would be detected solely by the detector $+1$,
and the detection time $t_{\E}$ would indicate the moment of emission.
In this case, for any local setting of the phase $\phi$, the
detections behind the left interferometer would either be coincident
with the detections on the right side at $t_E$ (we shall call this an
\emph{early} detection), or delayed at $t_{\L}=t_{\E}+\Delta L/c$ (a
\emph{late} detection).  This must be determined by the LHV model.
Half of the events on the left side are early (E) and half are late
(L). With the right interferometer in place, 1/4 of the events are
early on the left and late on the right (EL), 1/4 are late on the left
and early on the right (LE), and 1/2 are coincident. These coincident
events must then consist of equal parts early-early (EE) and late-late
(LL) events; no such distinction exists in the quantum description.

\begin{figure}[htbp]
  \begin{center}
    \psfig{file=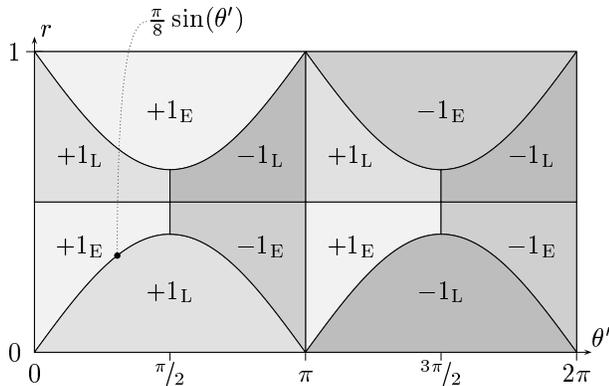} \caption{LHV model for detections at the
      left station.  The shifted value of the angular hidden variable,
      $\theta'=\theta-\phi$, and $r$, determine the result of the
      local observable, $l=\pm1$, and whether the particle is detected
      \emph{early} E or \emph{late} L. The lower curve in the left
      side of the chart is given by $\frac{\pi}{8}\sin{\theta'}$, and
      the shape of the other curves are of similar form.}
 \label{fig:det1}
  \end{center}
\end{figure}
 
\begin{figure}[htbp]
  \begin{center}
    \psfig{file=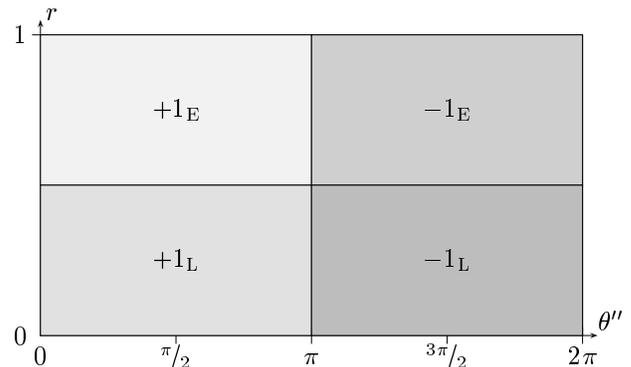} \caption{The measurement result at the
      right station given by the shifted hidden variables.  The
      symbols have the same meaning as in Fig.~\ref{fig:det1}.}
    \label{fig:det2}
  \end{center}
\end{figure}

In our model, the hidden variables are chosen to be an angular
coordinate $\theta\in[0,2\pi]$ and an additional coordinate
$r\in[0,1]$.  The ensemble of hidden variables is chosen as that of a
uniform distribution in this rectangle in $(\theta,r)$-space; each
pair of particles is then described by a definite point $(\theta,r)$
in the rectangle, defined at the source at the moment of emission. At
the left detector station, the measurement result is decided by the
hidden variables $(\theta,r)$ and the local setting $\phi$ of the
apparatus. When a photon arrives at the detection station, if the
interferometer works properly \cite{BLOCK} the variable $\theta$ is
shifted by the current setting of the local phase shifter (i.e.\ 
$\theta'=\theta-\phi$), and the result is read off
Fig.~\ref{fig:det1}.
At the right detector station, a similar procedure is followed
\cite{BLOCK,SYM}.  In this case, the shift is to the value
$\theta''=\theta+\psi$, and the result is obtained in
Fig.~\ref{fig:det2} in the same manner as before.

The single-particle detection probabilities straightforwardly follow
the quantum predictions, because in both Figs.~\ref{fig:det1}
and~\ref{fig:det2}, the total areas corresponding to $+1_{\E}$,
$-1_{\E}$, $+1_{\L}$, and $-1_{\L}$ are all equal. The particle is
equally likely to arrive early or late, and equally likely to go to
the $+1$ or $-1$ output port of the interferometer. The coincidence
probabilities are determined by interposing the two figures with the
proper shifts. 

\begin{figure}[htbp]
  \begin{center}
    \psfig{file=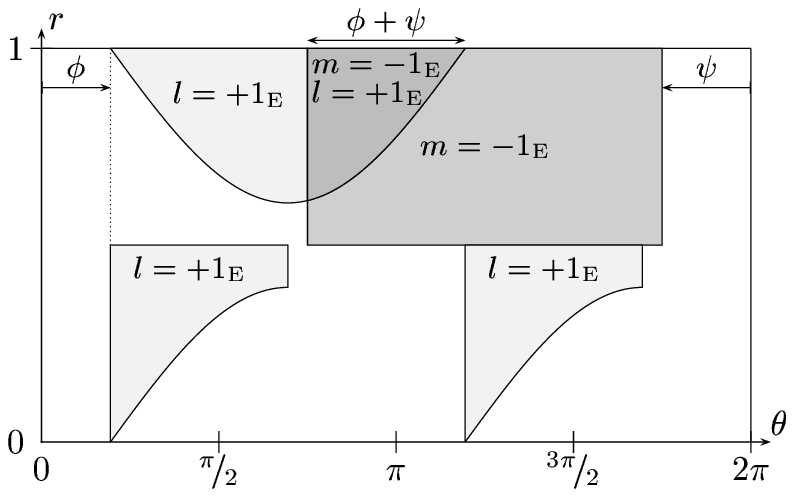} \caption{The shaded regions give the
      values for the initial hidden variables for which $l=+1_{\E}$ or
      $m=-1_{\E}$ are obtained (note that $\theta'=\theta-\phi$ while
      $\theta''=\theta+\psi$).  The overlap region of length
      $\phi+\psi$ represents the hidden variables for which both
      $l=+1_{\E}$ and $m=-1_{\E}$ are obtained.  }
    \label{fig:turn}
  \end{center}
\end{figure}

For example, the probability of having $l=+1_{\E}$ and $m=-1_{\E}$
simultaneously is the area of the set indicated in Fig.~\ref{fig:turn}
divided by $2\pi$ (the total area is $2\pi$ whereas the total
probability is 1). The net coincidence probability is
\begin{eqnarray}
  && P\bigl(+1;-1 (\text{coinc.})|\phi,\psi\bigr)\nonumber\\
  &&\qquad=P(+1_{\E};-1_{\E}|\phi,\psi)
  + P(+1_{\L};-1_{\L}|\phi,\psi)\nonumber\\
  &&\qquad =\frac{2}{2\pi}
  \int_0^{\phi+\psi}\frac{\pi}{8}\sin(\theta)d\theta
  =\frac{1}{8}\bigl(1-\cos{(\phi+\psi)}\bigr).
\end{eqnarray}
It is easy to verify that this model also gives the correct prediction
for the other detection events.

Somewhat remarkably, the above construction implies that \emph{the
  Franson experiment does not and cannot violate local realism if one
  disregards the fact that the unbalanced Mach-Zehnder interferometers
  are extended objects.} The reason that this construction is possible
is that the $50\%$ post-selection procedure discussed above may yield
\emph{an ensemble of detected pairs that depends on the phase
  settings} (rendering the Bell inequality useless \cite{ENSEMBLE}).
However, we shall now show that if the phase switching is performed at
the time-scale $\Delta L /c$, typical for retardations within the
interferometers, there is \emph{no} LHV description of the experiment.
In particular, we will describe an experimental procedure that allows
us to post-select an unchanging part of the LHV ensemble, thus
reenabling the Bell inequality on this part of the ensemble.

Let us look at one interferometer as an extended object to establish
what would take place if local realism were to hold. In the
interferometer, the decision of a detection to occur early (at
$t_{\E}$) or late (delayed by $\Delta L/c$) cannot be made later than
the time $t_{\E}$. This decision is based on the local variables and
the \emph{properly retarded} phase setting. No phase setting after
$t_{\E}-t_{\ret}$ can causally affect this E/L choice
\cite{RETARDATION}. The choice $\pm 1$ is also based on the local
variables and the properly retarded phase setting at the
interferometer in question, but this choice may be made as late as the
detection time $t_{\d}$ ($t_{\d}=t_{\E}$ for early events or
$t_{\d}=t_{\L}\equiv t_{\E}+\Delta L/c$ for late). Therefore, in the
case of a \emph{late} detection, the choice E/L and the choice $\pm 1$
can be made at different times ($t_{\E}$ and $t_{\L}$, respectively)
\emph{based on possibly different phase settings}.

Looking at only one interferometer, it is not possible to discern
early detections from late ones, so an experimenter at that
interferometer knows only the result $\pm 1$, the detection time
$t_{\d}$, and two \emph{possibly different} phase settings at
$t_{\d}-\Delta L/c-t_{\ret}$ and $t_{\d}-t_{\ret}$. She also knows
that for the events that are late, the later of these two phase
settings cannot causally have affected the E/L decision, so the
hypothetical late subensemble does not depend on the phase setting at
$t_{\d}-t_{\ret}$ but only on the phase setting at the earlier time
$t_{\d}-\Delta L/c-t_{\ret}$. By rejecting events where the phase
setting at $t_{\d}-\Delta L/c-t_{\ret}$ does not have a certain value
($\phi_0$, say), she ensures that the late subensemble does not change
at all.  To allow for settings other than $\phi_0$ at the later
decision time, a device which switches fast (on the time-scale $\Delta
L/c$) and randomly between phase settings is needed \cite{RANDOM}.

Thus, in the modified full experiment both experimenters should use
fast devices that randomly switch between the phase settings $\phi_0$,
$\phi_1$, $\ldots$, $\phi_N$ on the left side and $\psi_0$, $\psi_1$,
$\ldots$, $\psi_N$ on the right. They record the appropriate data and
reject (a) pairs of events whose registration times differ by $\Delta
L/c$ and (b) pairs of events which do not have the feature that
\emph{the phase setting at} $t_{\d}-\Delta L/c-t_{\ret}$ \emph{was
  $\phi_0$ on the left and $\psi_0$ on the right}.  The latter event
rejection ensures that the hypothetical LL subensemble within the
remaining data is independent of the phase settings at
$t_{\d}-t_{\ret}$.  Then, if local realism holds, the Bell-CHSH
inequality applies to this LL subensemble,
\begin{eqnarray}
   &&|\El(\phi_1,\psi_1)+\El(\phi_2,\psi_1)|\nonumber\\
   &&+|\El(\phi_2,\psi_2)-\El(\phi_1,\psi_2)| \le2,
\label{eq:CHSH}
\end{eqnarray}
where the phases are taken at $t_{\d}-t_{\ret}$, and $\El(\phi,\psi)$
denotes the Bell-type \emph{conditional} correlation function on the
remaining LL subensemble.  This is valid only because each of the
correlation functions above is an average on the same ensemble. Had
the ensemble depended on the phase settings at $t_{\d}-t_{\ret}$, the
bound would have been higher.

Indeed, the remaining EE subensemble may still depend on the phase
setting at $t_{\d}-t_{\ret}$ even after this selection, and we only
have
\begin{equation}
  |\Ee(\phi,\psi)|\le 1.  \label{eq:Eelimit}
\end{equation}
Experimentally, this ``noise'' in form of EE events cannot be
distinguished from the LL events.  Of all events that survive the
described selection, again half are EE and half are LL, so that
\begin{equation}
  \Es(\phi,\psi)
  =\frac12\El(\phi,\psi)
  +\frac12\Ee(\phi,\psi). \label{eq:halfhalf}
\end{equation}
Thus, a modified Bell-CHSH inequality valid for all the coincident
events is implied by (\ref{eq:CHSH})--(\ref{eq:halfhalf}), namely
\begin{eqnarray}
  &&|\Es(\phi_1,\psi_1)+\Es(\phi_2,\psi_1)|\nonumber\\
  &&+|\Es(\phi_2,\psi_2)-\Es(\phi_1,\psi_2)|
  \textstyle
  \le\frac12(2+4)=3.
\end{eqnarray}

Unfortunately, this inequality is not violated by the conditional
quantum correlation function
$\Es^{\text{QM}}(\phi,\psi)=\cos{(\phi+\psi)}$ which yields a maximum
of $2\sqrt{2}$. However, a violation may be obtained by a ``chained''
extension of the Bell-CHSH inequality (see Ref.\ \cite{CHAIN}):
\begin{eqnarray}
  &&|\El(\phi_1,\psi_1)+\El(\phi_2,\psi_1)|\nonumber\\
  &&+|\El(\phi_2,\psi_2)+\El(\phi_3,\psi_2)|\nonumber\\
  &&+|\El(\phi_3,\psi_3)-\El(\phi_1,\psi_3)|\le 4.
  \label{eq:chain}
\end{eqnarray}
If local realism holds, (\ref{eq:Eelimit}), (\ref{eq:halfhalf}), and
(\ref{eq:chain}) yield
\begin{eqnarray}
  &&|\Es(\phi_1,\psi_1)+\Es(\phi_2,\psi_1)|\nonumber\\
  &&+|\Es(\phi_2,\psi_2)+\Es(\phi_3,\psi_2)|\nonumber\\
  &&   +|\Es(\phi_3,\psi_3)-\Es(\phi_1,\psi_3)|\le
  \textstyle
  \frac{1}{2}(4+6) =5.
\label{eq:our}
\end{eqnarray}
This inequality \emph{is} violated by quantum predictions, e.g., at
$\phi_1=0$, $\phi_2=-\pi/3$, $\phi_3=-2\pi/3$, $\psi_1=\pi/6$,
$\psi_2=\pi/2$, and $\psi_3=5\pi/6$ we obtain
\begin{equation}
  5\cos(\pi/6)-\cos(5\pi/6)=
  6\cos(\pi/6)\approx5.20>5.
\end{equation}

In conclusion, to obtain a violation of local realism in an experiment
one needs random fast switching and a filtering of the hypothetical
late-late subensemble so that this ensemble does not depend on the
phase settings \cite{RANDOM}. Even then, the standard Bell
inequalities are not sensitive enough to show a violation of local
realism in the experiment, because their bound is raised by the
``noise'' introduced by the early-early subensemble. However, a
``chained Bell inequality'' may be used, which is violated even with
this ``noise'' included.

The reported violations of local realism from Franson experiments have
to be reexamined. While the results formally violate the standard
Bell-CHSH inequality, the inequality is not applicable. The inequality
(\ref{eq:our}) \emph{is} applicable, but when using it, one should
note that it is violated only if the visibility is more than
$5/5.2\approx 96\%$. This is significantly higher than the usual
$71\%$ bound discussed in the reported experiments
\cite{FR-EXP,CRYPTO}.

It has been proposed that entangled photons can be used to perform
quantum cryptography \cite{EKERT}; specifically, the Franson-type
experiment has been discussed in this context \cite{CRYPTO}. In such
schemes security checks can be performed by testing whether the
signals violate the Bell inequalities. It remains a subtle question if
the link to local realism is important for this kind of security
check; if so, the Bell-CHSH inequality is not appropriate for the
Franson setup.

Sven Aerts acknowledges a grant by the Flemish Institute for the
Advancement of Scientific-Technological Research in the Industry
(IWT). Jan-{\AA}ke Larsson has received partial support from the
Swedish Natural Science Research Council.  Marek \.Zukowski was
supported by the Flemish-Polish Scientific Collaboration Program No.\ 
007, by UG Program BW/5400-5-0264-9, and also acknowledges discussions
with E.\ Santos, H.\ Weinfurter and A.\ Zeilinger.


\begin{references}

 \bibitem[*]{SAERTSEMAIL}Electronic address: saerts@vub.ac.be
 \bibitem[\dagger]{KWIATEMAIL}Electronic address: kwiat@lanl.gov
 \bibitem[\ddagger]{JALAREMAIL}Electronic address: jalar@mai.liu.se
 \bibitem[\parsign]{FIZMZEMAIL}Electronic address: fizmz@univ.gda.pl
   
 \bibitem{FRANSON89} J. D. Franson, Phys. Rev. Lett. {\bf 62}, 2205
   (1989).  We are not interested here in modifications of Franson's
   idea like those in D. V. Strekalov, T. B. Pittman, A. V. Sergienko,
   Y. H. Shih and P. G.  Kwiat, Phys. Rev. A {\bf 54}, R1 (1996).
  
 \bibitem{FR-EXP} P. G. Kwiat, W. A. Vareka, C. K. Hong, H.  Nathel
   and R. Y. Chiao, Phys. Rev. A {\bf 41}, 2910 (1990); Z.  Y. Ou, X.
   Y.  Zou, L. J. Wang and L. Mandel, Phys. Rev.  Lett. {\bf 65}, 321
   (1990).  The first experiment with high time resolution is by J.
   Brendel, E. Mohler and W.  Martiensen, Phys. Rev. Lett. {\bf 66},
   1142 (1991), however their arrangement involved Michelson
   interferometers. The first full realization seems to be in P. G.
   Kwiat, A. M. Steinberg and R. Y.  Chiao, Phys. Rev. A {\bf 47},
   R2472 (1993).
  
 \bibitem{CRYPTO} P. R.  Tapster, J. G. Rarity, and P. C. M.  Owens,
   Phys. Rev. Lett., {\bf 73}, 1923 (1994), W.  Tittel, J.  Brendel,
   H.  Zbinden, and N.  Gisin, \emph{ibid.} {\bf 81}, 3563 (1998).
  
 \bibitem{CLASSICAL}J. D. Franson, Phys. Rev. Lett. {\bf 67}, 290
   (1991); Z. Y. Ou and L. Mandel, J. Opt. Soc. Am. B {\bf 7}, 2127
   (1990).
   
 \bibitem{GARUCCIO} L. De Caro and A. Garuccio, Phys.  Rev. A {\bf
     50}, R2803 (1994).
   
 \bibitem{KWIAT2} P. G. Kwiat, Phys. Rev. A {\bf 52}, 3380 (1995).
  
 \bibitem{SANTOS} The LHV model of E. Santos [Phys. Lett. A {\bf 212},
   10 (1996)] is limited to detection efficiency lower than $2/\pi$.
   
 \bibitem{MANDELSHIH} Z. Y. Ou and L. Mandel, Phys. Rev.  Lett.  {\bf
     61}, 50 (1988); Y. H. Shih and C. O. Alley, \emph{ibid.}  {\bf
     61} 2921 (1988).
  
 \bibitem{KWIAT} P. G. Kwiat, P.  H. Eberhard, A. M. Steinberg and R.
   Y.  Ciao, Phys. Rev. A {\bf 49}, 3209 (1994).
 
 \bibitem{POPESCU} S. Popescu, L. Hardy and M. \.Zukowski, Phys.  Rev.
   A., {\bf 56}, R4353 (1997).
  
 \bibitem{AD-HOC} The model does not take into account the fact that
   the local interferometer and the detection station are extended
   objects.  This is reasonable, provided the state of the
   interferometer does not undergo rapid changes at the time scale
   $\Delta L/c$.
  
 \bibitem{BLOCK} In case the interferometer is dismantled, the
   detection is always $+1_{\E}$. If one path is blocked, the events
   are randomly chosen from $+1$ or $-1$ each with probability 1/4
   (early or late as appropriate), or ``no detection'' with
   probability 1/2.  A modification of this type may be made as long
   as the assumption in \cite{AD-HOC} is valid.
  
 \bibitem{SYM} The model can be trivially symmetrized.
  
 \bibitem{ENSEMBLE} J.-{\AA}. Larsson, Phys. Rev. A, {\bf 57} 3304
   (1998).
  
 \bibitem{RETARDATION} The proper retardation for the E/L choice is
   really along the shortest phase-switch--detector path (not the
   optical path). It is possible to arrange the experiment so the two
   are roughly of the same size, and our argument still holds.
   
 \bibitem{RANDOM} Given that the switching is fast (on the time-scale
   $\Delta L/c$) and random, it is possible to prove the inequalities
   even without this filtering on the late ensemble.
   
 \bibitem{CHAIN} P. M. Pearle, Phys. Rev. D {\bf 2}, 1418 (1970); A.
   Garuccio and F. Selleri, Found. Phys. {\bf 10}, 209 (1980); S. L.
   Braunstein and C. M. Caves, Proc. 3rd Int. Symp. on Foundations of
   quantum mechanics, eds. S Kobayashi et al, (Phys.  Soc. Japan,
   Tokyo 1989).
  
 \bibitem{EKERT} A. K. Ekert (1991), Phys. Rev. Lett. {\bf 67}, 661.

\end{references}
\end{document}